# MULTIUSER DETECTION IN ASYNCHRONOUS MULTIBEAM COMMUNICATIONS

Helmi Chaouech[1] and Ridha Bouallegue[2]

[1]National Engineering School of Tunis, University of El-Manar, Tunis, Tunisia
`helmi.chaouech@planet.tn`
[1,2]Innov'COM Laboratory, Sup'COM, University of Carthage, Tunis, Tunisia
`ridha.bouallegue@gnet.tn`

## ABSTRACT

*This paper deals with multi-user detection techniques in asynchronous multibeam satellite communications. The proposed solutions are based on successive interference cancellation architecture (SIC) and channel decoding algorithms. The aim of these detection methods is to reduce the effect of co-channel interference due to co-frequency access, and consequently, improves the capacity of the mulitbeam communications systems, by improving frequency reuse. Channel estimation allows the determination of interference coefficients, which helps their effects compensation. The developed multi-user detections techniques are iterative. Therefore, detection quality is improved from a stage to another. Moreover, a signals combining method, which is integrated into these detection solutions, enhances their capability. The proposed solutions are evaluated through computer simulations, where an asynchronous multibeam satellite link is considered over an AWGN channel. The obtained simulation results showed the robustness of these multi-user detection techniques.*

## 1. INTRODUCTION

With the explosive evolution of information and communications technologies, mobility and flexibility of terminals become a necessity for several kinds of services. Wireless systems constitute a basic solution to satisfy these needs. In fact, they guarantee wireless and mobile communications for users independently of their localities. Thus, wireless communications solved some deficiencies of wired solutions. These contributions are mainly due to the transmission support; the atmosphere. The last allows data transmission every where; but an efficient physical resource sharing between users is necessary. Moreover, wireless transmissions are subject of several problems such as fading, noise, interference, multipath etc… These natural problems are dealt with robust signal processing techniques at the receivers.

As an example of wireless systems, satellite stations are good solutions which provide wide coverage and different communications services. These systems can guarantee network coverage in isolated places, where wired infrastructures and terrestrial wireless stations are difficult to install. Satellite systems offer also some specific services such as broadcasting, localisation, tracking etc.

Wireless channel is a common transmission support, which is shared between several simultaneous communications. Thus, efficient access and use of this communication mean is of great importance. Frequency is the main characteristic of wireless channels. Then, optimal division and reuse of this physical resource is needed to design high capacity communications systems. For wireless terrestrial networks, adjacent transceiver stations use different frequencies. In satellite systems, multibeam technology is a good solution for frequency reuse. This method forms at the satellite receiver a throng of narrow beams instead of a single wide beam. Each beam is defined by it carrier frequency. Thus, these different beams cover terrestrial zones with a co-frequency reuse for no adjacent cells. As a result, the capacity of the satellite system increases at the cost of co-channel interferences (CCI) and multiple access interference (MAI). The problems, which are due to these interferences and the noise, will be dealt by the receiver.

In this paper, we have developed some multi-user detection techniques for asynchronous multibeam systems. The aim of these solutions is to deal with co-channel interference, and provide an efficient frequency reuse. The proposed solutions are based on channel decoding, channel estimation, and interference cancellation, which operate in iterative processes. Due to asynchronous access of users to the system, propagation delay estimation task is needed as a first processed operation by the receiver. The developed techniques take advantage from the spatial diversity due to satellite antenna array. Thus, a signals combining solution, integrated in the multi-user detection methods, allows signal to noise ratio (SNR) improvement which leads to better detection quality.

The remainder of this paper is organized as follows. Section 2 presents related work. In section 3, the signal model for transmission and at the input of multi-user detection techniques is detailed. Section 4 shows the architecture of the system with its different functional blocks, and it expresses mathematically each block's operation. In section 5, simulations results are presented with some interpretations. At last, section 6 gives some conclusions from this work and presents propositions which can be subjects for the future works.

## 2. RELATED WORK

In this section, we present an overview of some related solutions. In fact, multi-user detection and channel estimation are research fields of big importance, especially for wireless communications. Thus, in the last three decades, many authors and several works are concentrated on detection and channel estimation techniques. Some of them, dealt with joint detection and channel estimation problems [1], [2], [3]. In fact, detection techniques need channel effects compensation in order to can combat noise and multiple access interference. In [4] and [5], some channel estimation techniques and propagation delay estimation methods are presented and evaluated. Some wireless channel models are discussed and evaluated in [6]. In [7], a channel estimation solution for mulibeam communication is developed. In [8], a channel estimation technique combined with a multi-user detection method is developed and evaluated. Evaluation of detection techniques can be done by bit error rate (BER) computation, and by their implementation complexity, as their processor time consummation. In his book [9], Verdu developed, analyzed and evaluated some multi-user detection techniques. Thus, based on their mathematical formulations, detection solutions can be classified into some categories. The conventional detection or the matched filtering detector suffers from co-channel interference and is very sensitive to near-far problem [9], [10]. The maximum likelihood (ML) detection technique, which is known as the optimal detector, is developed to solve optimally the weakness of the conventional detection [11]. This ML based detector has optimal performance in the presence of MAI and near-far problem at the cost of computational complexity, which does not promote its practical implementation particularly for real-time applications. The linear multi-user detectors, which are the decorrelating technique and the minimum mean-squared error (MMSE) detector, deal with MAI robustly, and they are near-far resistant [9], [10], [12], [13]. However, their major weaknesses are noise enhancement and inversion of big dimension matrices, especially in the case of asynchronous communications. Iterative interference cancellation detection techniques are robust solutions which subtract CCI and improve the BER iteratively [14], [15], [16], [17]. With these detection techniques, signals are cleaned of MAI successively or in a parallel processing. Thereafter, they can be classified in two main architectures; the successive interference cancellation (SIC) methods and the parallel interference cancellation (PIC) ones. Coded multi-user detections showed high performance. That's why; they occupy a big amount of the works which deal with detection problems. In fact, this kind of detection solutions, in addition to its interference cancellation, relies on channel decoders to combat noise and interference. These multi-user detection techniques give acceptable BERs even with low SNRs. Some detection techniques incorporating channel decoding are presented in [18], [19], [20], [21], [22]. Other detection solutions developed for multibeam communications, which are detailed and evaluated, can be found in [23], [24].

## 3. SIGNAL MODEL

We consider the asynchronous uplink of a multibeam satellite system. K active users share a co-frequency channel and send their signals to satellite antenna array. The K users belong to different cells which are covered by co-frequency beams. Users in the same cell adopt a TDMA access to physical resources. Thus, co-channel interferences are due to same frequency reuse. Mathematically, we represent the k$^{th}$ signal by:

$$r_k(t) = a_k e^{j\varphi_k} x_k(t) \tag{1}$$

Where, $a_k$ and $\varphi_k$ are the amplitude of the signal and its carrier phase, and $x_k(t)$ is given by:

$$x_k(t) = \sum_{i=0}^{N-1} x_k[i] g(t - iT - \tau_k) \tag{2}$$

With, $x_k[i]$, $i = 0..N-1$ is a sequence of N QPSK symbols, $g(t)$ is the emitter filter waveform, $T$ is the symbol temporal duration and $\tau_k$ is the propagation delay of the k$^{th}$ signal. For simplicity and without loss of generality, we assume an ordering on the time delays such that: $\tau_1 \le \tau_2 \le \ldots \le \tau_K < T$. We suppose that the propagation delays are multiple of $T/N_s$; the sampling period. Thus, $N_s$ represents the sampling factor or the number of samples taken by symbol during. The sequence of $N$ symbols, which are convolutional coded and interleaved before transmission, is divided into $N_p$ pilot symbols and $N_i$ information symbols.

This signal, expressed in (2), is received by the L radiating components of the antenna array, (see figure 1). Thus, if we generalize for the K users, the signal received by the $l^{th}$ element of the antenna can be expressed by:

$$s_l(t) = \sum_{k=1}^{K} d_k^{(l)} r_k(t) + n_l(t) \tag{3}$$

With, $d_k^{(l)}$ is the $l^{th}$ coefficient of the steering vector $d_k$ of the k$^{th}$ received signal, and $n_l(t)$ is a Gaussian noise, added to the composite signal received by the $l^{th}$ antenna component.

To arrange the L signals received by the antenna array sensors, we define the signal vector $s(t)$ by: $s(t) = [s_1(t), s_2(t), \ldots, s_L(t)]^T$. Where, $(.)^T$ denotes the transpose operator. Thus, generalization of equation (3) for the L components of the antenna gives:

$$s(t) = \sum_{k=1}^{K} d_k r_k(t) + n(t) \tag{4}$$

Where, $d_k$, as it is mentioned above, is the column vector of length L which contains information about the arrival direction of the k[th] signal [25], and $n(t) = [n_1(t), \ldots, n_L(t)]^T$ is the additive noise vector at the L radiating elements outputs. We define the direction of arrival (DOA) matrix of the K beams by: $D = [d_1, \ldots, d_K]$, and the vector of received signals: $r(t) = [r_1(t), \ldots, r_K(t)]^T$. Thus, equation (4) can be rewritten as follows:

$$s(t) = Dr(t) + n(t) \qquad (5)$$

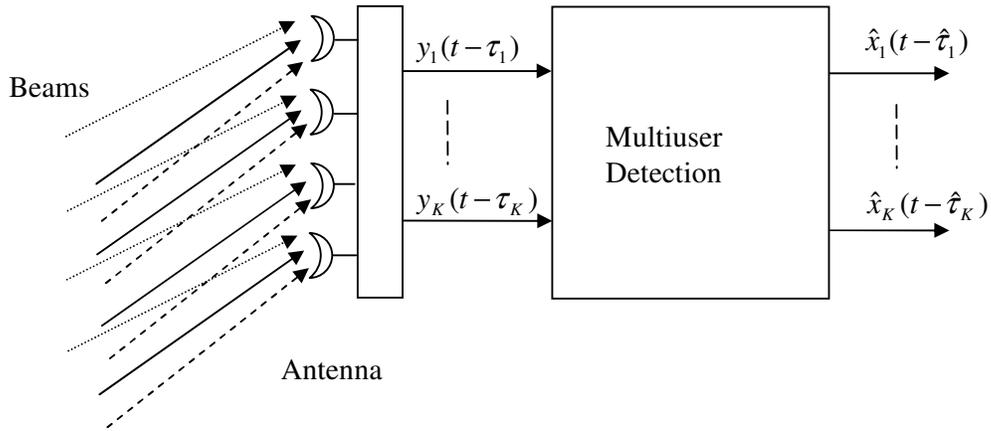

Figure 1. Reception model.

At the receiver, beam forming operation maximizes the energy of the useful signal by steering the antenna array in the DOA of this beam [25]. Thereafter, to form the k[th] beam, the treatment consists of taking a linear combination of the signals at the antenna elements outputs. The obtained signal for the k[th] beam is given by:

$$y_k(t) = v_k^{\,T} \, s(t) \qquad (6)$$

With, $v_k$ is a column vector which contains the L coefficients of the k[th] beam forming. Generalization of the expression (6) for the K active users in the system gives:

$$y(t) = Vs(t) \qquad (7)$$

With, $y(t) = [y_1(t), y_2(t), \ldots, y_K(t)]^T$ and $V = [v_1, v_2, \ldots, v_K]^T$.

Using (5), equation (7) becomes:

$$y(t) = VDr(t) + Vn(t) = Wr(t) + Vn(t) \qquad (8)$$

We define the diagonal matrix of the K signals complex amplitudes by: $A = diag\left(\left[a_1 e^{j\varphi_1}, \ldots, a_K e^{j\varphi_K}\right]\right)$. With $diag(.)$ denotes the matrix diagonal operator. Using the expression of $r(t)$ in (1), equation (8) can be rewritten also:

$$y(t) = WAx(t) + Vn(t) \qquad (9)$$

Where, $x(t) = [x_1(t), \ldots, x_K(t)]^T$

By introducing the channel matrix $H = WA$, and the noise vector $z(t) = Vn(t)$, equation (9) becomes:

$$y(t) = Hx(t) + z(t) \qquad (10)$$

The discreet model, which is derived from (10) after optimal sampling, is given by:

$$y[i] = Hx[i] + z[i] \qquad (11)$$

Thus, these samples will be dealt by the multi-user detection technique in order to determine the original data sent by user's terminals.

## 4. MULTI-USER DETECTION

The proposed multi-user detection techniques are composed of some functional blocks which operate the following jobs: propagation delay estimation, phase estimation, channel decoding, channel estimation, interference cancellation and signals combining. These operations are executed successively. In each block of the receiver, signals users are also dealt successively. Some of these operations are processed iteratively in some stages and the others are dealt only at the beginning of algorithms running.

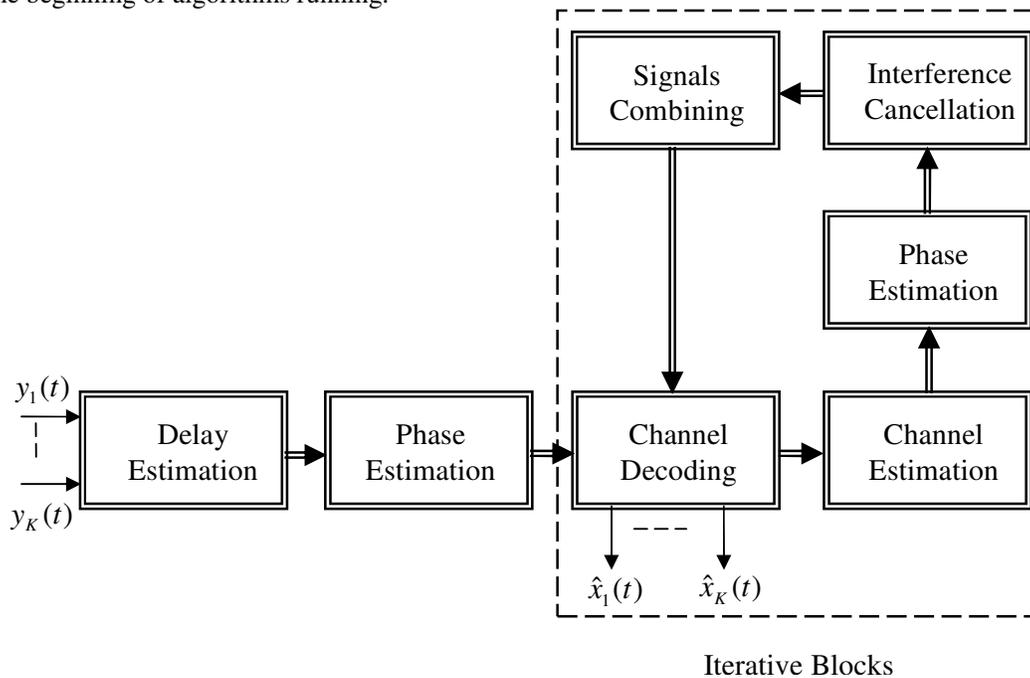

Iterative Blocks

Figure 2. Receiver architecture.

### 4.1. Delay Estimation

The first operation to be dealt is the propagation delays estimation. The results of this operation are then used by the following receiver blocks. Thus, the performance of the multi-user detection is strongly influenced by the delays estimation performances. In each processing operation of the receiver, the asynchronous character of the signals is not modified but taken into account. The propagation delay estimation technique is based on signals correlation, and it uses the pilot symbols. For the $k^{th}$ signal, estimation of its propagation delay can be expressed by:

$$\hat{\tau}_k = \arg Max\left(\left|\sum_{i=1}^{N_s \times N_p} \text{Re}\left(y_k(i).P_k(i)^*\right)\right| + \left|\sum_{i=1}^{N_s \times N_p} \text{Im}\left(y_k(i).P_k(i)^*\right)\right|\right)$$ (12)

Where, $\text{Re}(.)$ and $\text{Im}(.)$ denote the real part and imaginary part of a complex number respectively, and $(.)^*$ denotes the conjugate operator of complex number. $N_p$ is the pilot symbols number.

### 4.2. Phase Estimation

From figure 3, we note that the receiver contains two phase estimation blocks. In fact, these two operations are done differently. The first is an initialisation of the iterative algorithm. It is needed to compensate the phase effects before decoding algorithms processing. In the first stage of the multi-user detection techniques, an initial phase estimation of the $k^{th}$ signal is given by:

$$\hat{\varphi}_k = ang\left(\sum_{N_p} y_k(j)\tilde{P}_k(j)^*\right)$$ (13)

Where, $\tilde{P}_k$ is a $N_s \times N_p$ length column vector. It is derived from the pilot sequence vector $P_k$ as follows: $\tilde{P}_k = \left[P_k(1),..,P_k(1),\ldots,P_k(N_p),..,P_k(N_p)\right]^T$, with, $P_k$ is the vector of $k^{th}$ user training sequence which consists of $N_p$ pilot symbols. And, $ang(.)$ denotes the angle of a complex number operator.

The second phase estimation block, which is included in the iterative part of the algorithms, is performed after channel coefficients estimation. This iterative operation can be expressed, for the $k^{th}$ signal at the $n^{th}$ stage, as:

$$\hat{\varphi}_k^{(n)} = ang\left(\hat{h}_{k,k}^{(n)}\right)$$ (14)

With, $\hat{h}_{k,k}^{(n)}$ is the $(k, k)^{th}$ channel matrix estimation at the $n^{th}$ iteration. It is explained in section 4.4.

### 4.3. Channel Decoding

Before channel decoding, the signals phases are compensated with use of phases estimations. This operation consists to multiply the symbols samples by the quantities $e^{-j\hat{\varphi}_k^{(n)}}$ for the K users respectively. We have implemented two different decoding methods for the convolutional channel coding. They are the Viterbi algorithm [26], [27], and the BCJR one [28], [29]. The second technique needs SNR knowledge. Thus, SNR estimation operation, which is not presented in the above receiver architecture, is necessary. This task is explained in the following paragraph.

### 4.3.1. SNR estimation

The SNR estimation is computed in each iteration before BCJR or MAP (Maximum a Posteriori) decoding operation. It is based on the BCJR algorithm itself. That solution provides an estimation of the signal to noise ratio by minimizing the bit error rate between the pilot symbols and the decoded samples which correspond to the transmission of that pilot sequence. Thus, SNR estimation can be described by:

$$\hat{SNR} = \arg\left(\min_{snr \in [snr_{min} \ snr_{max}]} (BER)\right) \tag{15}$$

Other solutions of SNR estimation techniques can be found in [30].

### 4.4. Channel Estimation

The channel estimation technique allows the determination of CCI coefficients in order to compensate their effect in interference cancellation block. These coefficients, which are the channel matrix $H$ elements, are estimated iteratively. In each stage, they are updated, with use of estimated and pilot symbols. Channel coefficients estimation of the $k^{th}$ signal at the $n^{th}$ iteration can be expressed by the following equation:

$$\hat{h}_{k,l}^{(n)} = \frac{1}{2 \times N_s (N+1)} \sum_{j=1}^{N_s(N+1)} y_k(j)\hat{x}_l^{(n)}(j) \tag{16}$$

Where,

- $\hat{h}_{k,l}^{(n)}$ is the estimation of interference coefficient of signal $l$ on signal k at $n^{th}$ iteration, or the $(k,l)^{th}$ element estimation of the matrix $H$ at the at the $n^{th}$ iteration.
- The value 2 in the denominator is to compensate the effect of the square of the QPSK symbols modulus on the channel coefficients estimation.
- $\hat{x}_l^{(n)}(j)$ is the estimation of the $j^{th}$ symbol of user $l$ at the $n^{th}$ iteration.

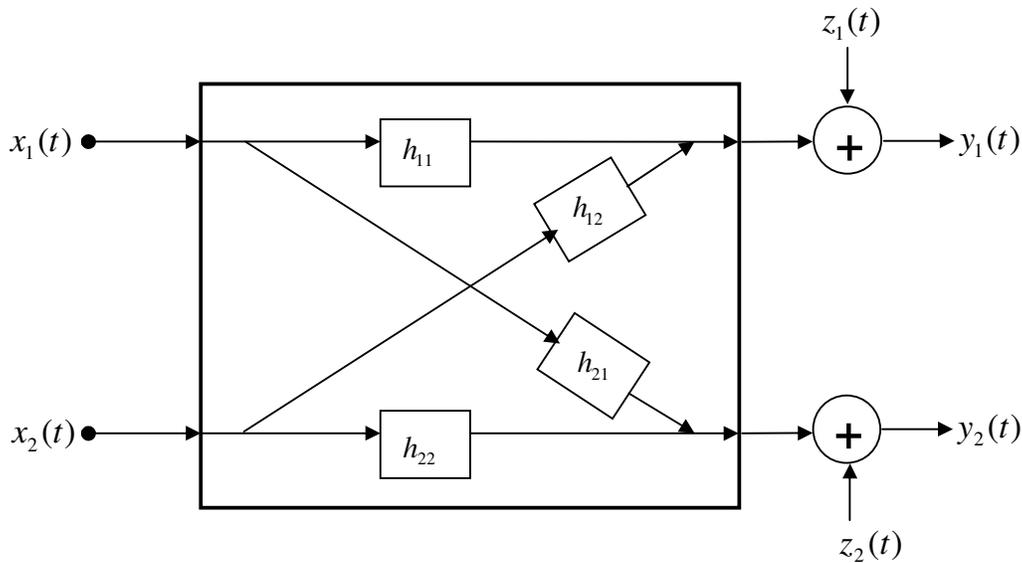

Figure 3. CCI Channel Model for K=2.

**4.5. Interference Cancellation**

The successive interference cancellation block ensures the CCI suppression from the K beams. Thus, co-frequency signals are dealt successively in each iteration. Symbols of interfering signals are extracted from the useful signal with use of the appropriate estimated channel coefficients, and taking into account of signals propagation delays. Thus, interference cancellation from the j[th] sample of k[th] signal at the n[th] iteration is expressed by:

$$y_k^{(n)}(j) = y_k(j) - \sum_{k=1}^{k-1} \hat{h}_k^{(n-1)}(k') x_k^{(n)}(j) - \sum_{k=k+1}^{K} \hat{h}_k^{(n-1)}(k') x_k^{(n-1)}(j) \tag{17}$$

**4.6. Signals Combining**

The technique of signals combining takes advantage from the spatial diversity due to the antenna array. This solution improves the SNR of the useful signal by extracting its interfering parts on the other beams, and adds them coherently to this signal. In this way, that method leads to better detection quality. Thus, for the j[th] sample of k[th] signal at the n[th] stage of the multi-user detection techniques, signals combining operation can be expressed by the two following equations:

$$\chi_k^{(n)}(j) = \sum_{\substack{k=1 \\ k' \neq k}}^{K} \hat{h}_k^{(n-1)}(k)^* \left( y_{k'}(j) - \sum_{\substack{k'=1 \\ k'' \neq k'}}^{K} \hat{h}_k^{(n-1)}(k'') \hat{x}_{k''}^{(n-1)}(j) \right) \tag{18}$$

$$\tilde{y}_k^{(n)}(j) = y_k^{(n)}(j) + \chi_k^{(n)}(j) \tag{19}$$

Thereafter, at the n[th] iteration, for the j[th] symbol of the k[th] signal, the decoder computes the quantity $\tilde{y}_k^{(n)}(j)$.

# 5. SIMULATIONS RESULTS

In order to evaluate the performances of the multi-user detection techniques, we have considered an asynchronous multibeam satellite reverse link scenario. Channel coding and modulation are taken those of the DVB RCS standard [31]. Thus, we have employed the convolutional coding of this system to code the transmission data frame. This convolutional code is defined by its two generator polynomials, which are represented by their octal form as: [171 133]. The modulation is QPSK. Before transmission, data is randomly interleaved. The other parameters of simulation, which characterize the applied scenario, are: K=5, Ni=100, Np=30 and Ns=4. We have performed the multi-user detection techniques with Monte Carlo simulations. In each iteration, N QPSK symbols and K carrier phases are randomly generated. The phases are uniformly distributed in $[0, 2\pi]$. The propagation delays, which belong to the temporal interval $[0, T[$, are also randomly generated. The signals amplitudes are taken equal, with unit powers; $a_1 = a_2 = \ldots = a_5 = 1$. We considered that the interferences are uniformly distributed between the co-frequency signals. Then, the CCI coefficients are of the same modulus. Thus, the CCI channel matrix can be presented by:

$$|H| = \begin{bmatrix} 1 & \mu & \mu & \mu & \mu \\ \mu & 1 & \mu & \mu & \mu \\ \mu & \mu & 1 & \mu & \mu \\ \mu & \mu & \mu & 1 & \mu \\ \mu & \mu & \mu & \mu & 1 \end{bmatrix} \qquad (20)$$

Where, $\mu = 0.25$.

The following figures show the simulations results. In order to evaluate the performances of the multiuser detection techniques, we have plotted in figures 4 and 5 the average BER evolution with different values of SNRs. The propagation delays estimation method is evaluated in figure 6, by computing delay estimation errors vs SNRs. In figure 4, the simulations results of the multi-user detection technique with Viterbi channel decoding are presented. The performances of this solution are good even with low SNRs. as an iterative algorithm, the detection is improved from a stage to another of the SIC multi-user architecture. We have shown, in figure 5, the simulation results of the detection technique implementing BCRJ channel decoding. Compared to the first solution, the obtained results, with MAP decoding, are better. But, when we ran algorithms via computer simulations, we noted that the viterbi decoding based technique is more than three once faster than the solution implementing BCJR algorithm. Thus, MAP decoding introduced, with the significant improvement performances, some complexity of processing at the receiver. Moreover, SNR estimation operation, which is performed before BCJR, decoding, consumed some processing time. If we look again on simulations results shown in figures 4 and 5, it is clear that signals combining technique enhanced the detection quality for both solutions. This operation allowed somehow the decreasing of stages number of the algorithms by converging rapidly to the desired bit error rates. For example, according to figures 4 and 5, stages 2, where signals combining techniques are applied, gave better results than stages 3 without signals combining application. In figure 6, we have evaluated the propagation delays estimation technique. Average error of delays estimation of the K users is presented for different values of the SNRs and CCI powers. The obtained results show the robustness of the proposed solution under a highly noisy environment, although some degradation of estimation quality, which is due to co-channel interference power increasing. Thus, high estimation quality of the propagation delays was of great importance for the other detection blocks, since the signals are dealt with asynchronously, without need of their synchronization as a prior task.

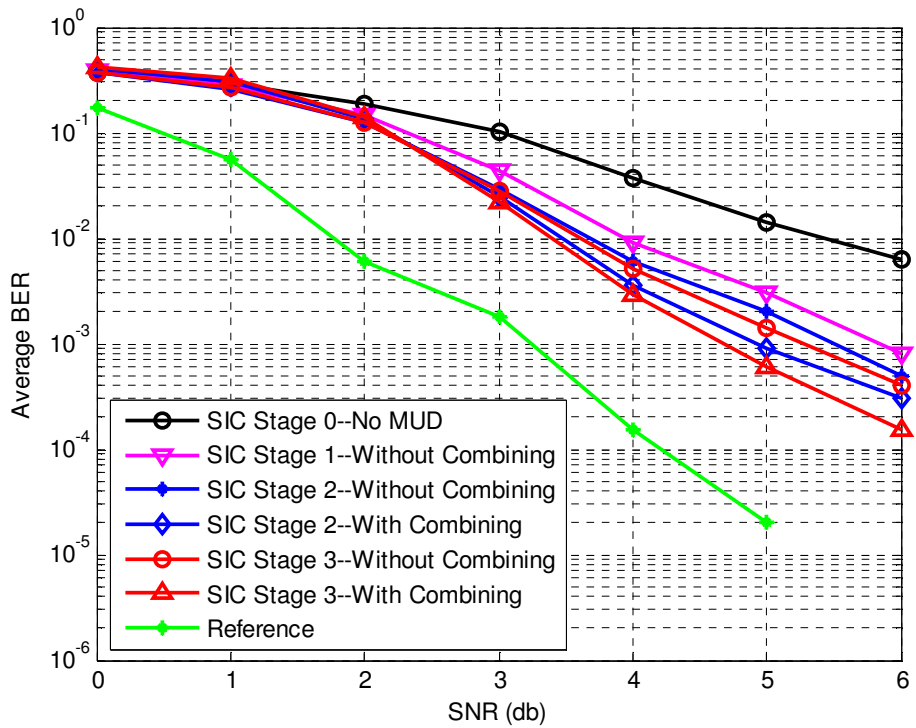

Figure 4. Performance evaluation of the detection with Viterbi decoding.

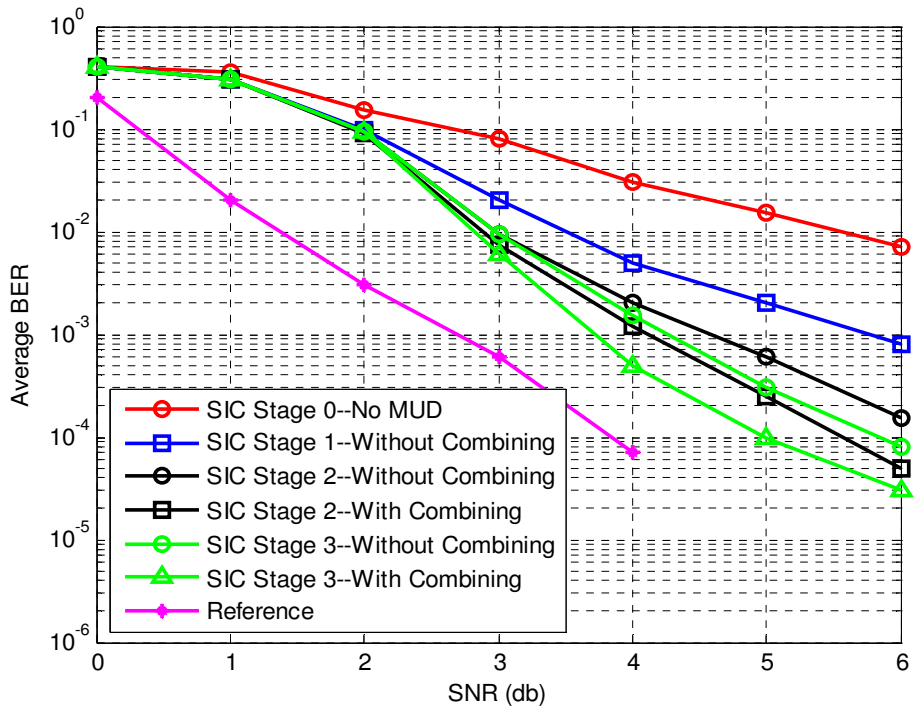

Figure 5. Performance evaluation of the detection with BCJR decoding.

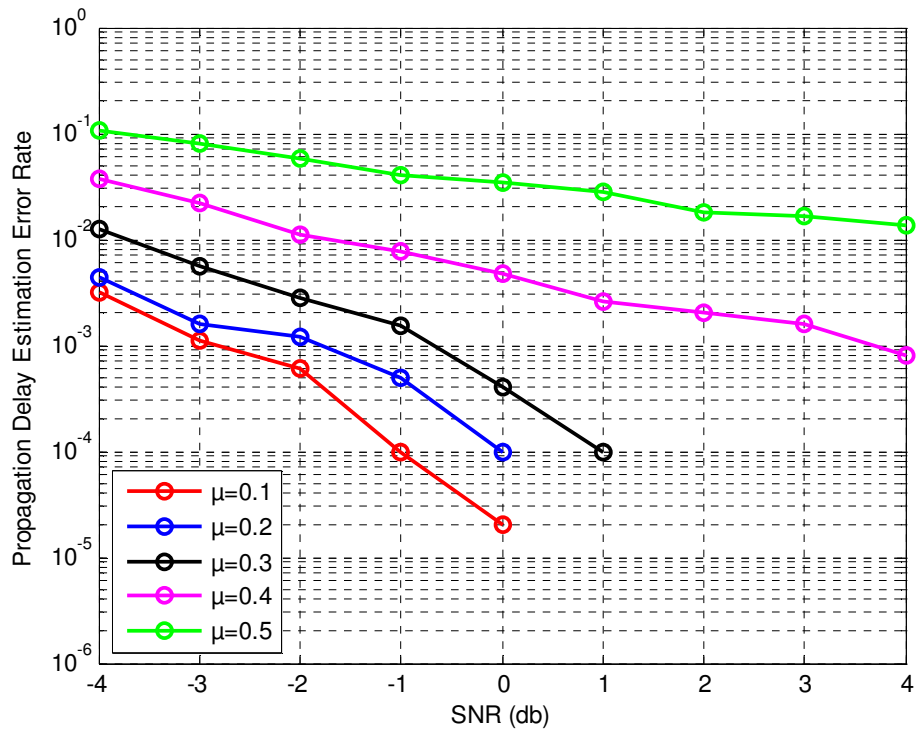

Figure 6.  Performance evaluation of the propagation delay estimation.

## 6. CONCLUSIONS

In this paper, we have developed and evaluated through computer simulations some multi-user detection solutions for asynchronous multibeam systems. The proposed techniques are based on successive interference cancellation architectures which implement channel decoding and estimation methods. We integrated two convolutional decoding algorithms. The Viterbi technique and the BCJR one. Both multi-user detection techniques showed good performances under noisy and CCI situation. Moreover, the quality of symbols detection is improved from a stage to another. The solution, which implements MAP decoding, gave better results compared to the detection technique employing Viterbi decoding algorithm. But, it introduced more processing complexity in the receiver due to probabilities calculation, in addition to the SNR estimation block, which is needed for MAP decoding. Thus, the multi-user solution, which integrates Viterbi algorithm, is faster in execution, than the other. As we dealt with asynchronous communications, estimation of propagation delays operation was of significant importance.  Its good performances helped enormously detection techniques processing. Through the simulations results, we can conclude the importance of the signals combining technique. This operation allowed SNR improvement of the useful signal, which leads to a better quality of detection, and therefore, a possible reduction of the stages number. As a future work, it will be interesting to deal with multi-user detection techniques in multibeam communications under an asynchronous multipath channel.

# REFERENCES

[1]     K. J. Kim and R.A. Iltis, "Joint detection and channel estimation algorithms for QS-CDMA signals over time-varying channels," IEEE Transactions on communications, Vol. 50, No. 5, May 2002.

[2]     T. Zemen, C. F. Mecklenbrauker, J. Wehenger and R.R. Muller, "Iterative joint time-variant channel estimation and multi-user detection for MC-CDMA," IEEE Transactions on wireless communications, Vol. 5, No. 6, June 2006.

[3]     S. Y. Park, B. Seo and C. Kang, "Performance of iterative receiver for joint detection and channel estimation in SDM/OFDM systems," IEICE Transactions on communications, Vol. E86-B, No. 3, March 2003.

[4]     M. Sirbu,  "Channel and delay estimation algorithms for wireless communication systems," Thesis of Helsinki University of Technology (Espoo, Finland), December 2003.

[5]     L. M. Davis, I. B. Collings and R. J. Evans, " Estimation of LEO satellite channels," International conference on information, communications and signal processing, pp. 15-19, Singapore, September 1997.

[6]     A. Amer and F. Gebali, "General model for infrastructure multichannel wireless LANs," International Journal of Computer Networks & Sommunications (IJCNC), Vol. 2, No. 3, May 2010.

[7]     H. Chaouech and R. Bouallegue, "Channel estimation and detection for multibeam satellite communications," IEEE Asia pacific conference on circuits and systems, pp. 366-369, Kuala Lumpur, Malaysia, December 2010.

[8]     H. Chaouech and R. Bouallegue, "Channel estimation and  multiuser detection in asynchronous satellite communications," International Journal of Wireless & Mobile Networks, Vol. 2, No. 4, pp. 126-139, November 2010.

[9]     S. Verdu, Multiuser Detection, Cambridge University Press, 1998.

[10]    K. Khairnar and S. Nema, "Comparison of Multi-user detectors of DS-CDMA System," World Academy of Science, Engineering and Technology, 2005.

[11]    S. Verdu, "Minimum probability of error for asynchronous gaussian multiple access channels," IEEE Transactions on information  theory, Vol. IT-32, No. 1, January 1986.

[12]    R. Lupas and S. Verdu, "Linear multi-user detectors for synchronous code-division multiple access channels," IEEE Transactions on information theory, Vol. 35, No. 1, pp. 123-136, January 1989.

[13]    R. Lupas and S. Verdu, "Near-far resistance of multiuser detectors in asynchronous channels," IEEE Transactions on communications, Vol. 38, No. 4, pp. 496-508, April 1990.

[14]    M. K. Varanasi, "Multistage detection in asynchronous code-division multiple-access communications," IEEE Transactions on communications, Vol. 38, No. 4, pp. 509-519, April 1990.

[14]    A. L. C. Hui and K. Ben Letaief, "Successive interference cancellation for multi-user asynchronous DS/CDMA detectors in multipath fading links," IEEE Transactions on communications, Vol. 46, No. 3, pp. 384-391, March 1998.

[16]    K. Ko, M. Joo, H. Lee, and D. Hong, "Performance analysis for multistage interference cancellers in asynchronous DS-CDMA Systems," IEEE Communications letters, Vol. 6, No. 12, December 2002.

[17]    S. H. Han and J. H. Lee, "Multi-stage partial parallel interference cancellation receivers for multi-rate DS-CDMA system," IEICE Transactions on communications, Vol. E86-B, No. 1, January 2003.

[18]    J. Hagenauer, E. Offer, and L. Papke, "Iterative decoding of binary block and convolutional codes," IEEE Transactions on information theory, Vol. 42, No. 2, pp.429-445, March 1996.

[19]    M. L. Moher, "An iterative multiuser decoder for near-capacity communications," IEEE Transactions on communications, Vol. 46, No. 7, pp. 970-880, July 1998.

[20]    M. L. Moher, "An iterative algorithm for asynchronous coded multi-user detection," IEEE Communications letters, Vol. 2, No. 8, pp. 229-231, August 1998.

[21]    H. H. Chen and Z. Q. Liu, "A CDMA multi-user detector with block channel coding and its performance analysis under multiple access interference," IEICE Transactions on communications letters, Vol. E81-B, No. 5, pp. 1095-1101, May 1998.

[22]    C. Schlegel, P. Alexander and S. Roy, "Coded asynchronous CDMA and its efficient detection," IEEE Transactions on information theory, Vol. 44, No. 7, pp. 2837-2847, November 1998.

[23]    M. L. Moher, "Multiuser decoding for multibeam systems," IEEE Transactions on vehicular technology, Vol. 49, No. 4, pp. 1226-1234, July 2000.

[24]    M. Debbah, G. Gallinaro, R. Muller, R. Rinaldo, and A. Vernucci, "Interference mitigation for the reverse- link of interactive satellite networks," 9th International workshop on signal processing for space communications (SPSC), ESTEC, Noordwijk, The Netherlands, 11-13 September 2006.

[25]    Lal. C. Godara, "Application of antenna arrays to mobile communications, Part II: Beam-forming and direction-of-arrival considerations," Proceedings of the IEEE, Vol. 85, No. 8, August 1997.

[26]    A. J. Viterbi, "Error bounds for convolutional codes and an asymptotically optimum decoding algorithm," IEEE Transactions on information theory, Vol. 13, pp. 260-269, April 1967.

[27]    G. D. Forney, "The Viterbi algorithm," Proceedings of the IEEE, Vol. 61, No. 3, pp.268-278, March 1973.

[28]    L. R. Bahl, J. Cocke, F. Jelinek, and J. Raviv, "Optimal decoding of linear codes for minimizing symbol error rate," IEEE Transactions on information theory, Vol. IT-20, pp. 284-287, March 1974.

[29]    P. Robertson, P. Villebrun, and P. Hoeher, "A comparison of optimal and sub-optimal MAP decoding algorithms operating in the log domain," IEEE International conference on communications, Seattle, Washington, June 1995, pp. 1009-1013.

[30]    D. R. Pauluzzi and N. C. Beaulieu, "A comparison of SNR estimation techniques for the AWGN channel," IEEE Transactions on communications, Vol. 48, pp. 1681-1691, October 2000.

[31]    ETSI EN 301 790 V1.3.1 (2003-03), "Digital Video Brodcasting (DVB); interaction channel for satellite distribution systems".

**Authors**

**Helmi CHAOUECH** received the engineering degree in telecommunications in 2006 and the M.S degree in communications systems in 2007 from the National Engineering School of Tunis, Tunisia. Currently, he is with the Innov'COM Laboratory at Higher School of Communications of Tunis (Sup'COM), Tunisia as a Ph.D student and he is an assistant in the Faculty of Economic Sciences and Management of Nabeul, Tunisia. His research interests include channel estimation, multi-user detection, CDMA systems, wireless communication theory and multibeam satellite communications.
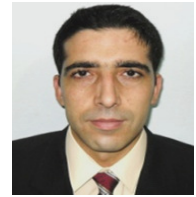

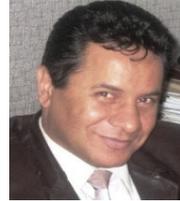

**Ridha BOUALLEGUE** received the Ph.D degree in electronic engineering from the National Engineering School of Tunis in 1998. In March 2003, he received the HDR degree in multi-user detection in wireless communications. From 2005 to 2008, he was the Director of the National Engineering School of Sousse, Tunisia. In 2006, he was a member of the national committee of science technology. Since 2005, he was the director of the 6'TEL Research Unit at Sup'COM. Since 2011, he was the founder and the director of Innov'COM Laboratory at Sup'com. Currently, he is the director of Higher School of Technology and Informatic, Tunis, Tunisia. His current rsearch interests include wireless and mobile communications, OFDM, space- time processing for wireless systems, multi-user detection, wireless multimedia communication and CDMA systems.